\documentclass[conference]{IEEEtran}
\IEEEoverridecommandlockouts

\usepackage{cite}
\usepackage{amsmath,amssymb,amsfonts}
\usepackage{algorithmic}
\usepackage{graphicx}
\usepackage{textcomp}
\usepackage{xcolor}
\usepackage{comment}
\usepackage{subcaption}
\usepackage{booktabs,multirow}
\usepackage{makecell}
\usepackage{url}
\usepackage{graphicx}
\usepackage{amsmath}
\def\BibTeX{{\rm B\kern-.05em{\sc i\kern-.025em b}\kern-.08em
    T\kern-.1667em\lower.7ex\hbox{E}\kern-.125emX}}
\begin{document}

\title{Rethinking Music Tokenization: A Semantic Codec toward High-Fidelity LLM
Music Generation\\

\author{
 \IEEEauthorblockN{
  Huakang Chen\IEEEauthorrefmark{2}, 
  Guobin Ma\IEEEauthorrefmark{2}, 
  Yuepeng Jiang\IEEEauthorrefmark{2}, 
  Dake Guo\IEEEauthorrefmark{2},
  Jingbin Hu\IEEEauthorrefmark{2},
  Hanke Xie\IEEEauthorrefmark{2},\\
  Wenhao Li\IEEEauthorrefmark{2}, 
  Lingxin Xiong\IEEEauthorrefmark{3}, 
  Jian Zhao\IEEEauthorrefmark{3}, 
  Zhonglin Jiang\IEEEauthorrefmark{3}, 
  Yong Chen\IEEEauthorrefmark{3}, 
  Lei Xie\IEEEauthorrefmark{2}, 
  Pengcheng Zhu\textsuperscript{*}\IEEEauthorrefmark{4}}\thanks{* Corresponding author.}
 \IEEEauthorblockA{\IEEEauthorrefmark{2}Audio, Speech and Language Processing Lab (ASLP@NPU), \\ School of Software, Northwestern Polytechnical University, Xi’an, China}
 \IEEEauthorblockA{\IEEEauthorrefmark{3}Geely Automobile Research Institute (Ningbo) Company Ltd, Ningbo, China} 
 \IEEEauthorblockA{\IEEEauthorrefmark{4}WeNet Open Source Community} 
 \IEEEauthorblockA{huakang@mail.nwpu.edu.cn, zpcoftts@gmail.com}
}

}

\maketitle

\begin{abstract}
Discrete audio tokenization has become the critical interface between raw waveforms and autoregressive modeling in recent music generation. As a result, music tokenizers must simultaneously support high-fidelity reconstruction and produce discrete sequences that remain amenable to language modeling. Existing reconstruction-oriented tokenizers often mix musical structure with fine acoustic details, producing high-entropy tokens that are hard to model. In contrast, semantics-guided alternatives are designed for speech and do not fit music well, often hurting reconstruction quality.
We address these trade-offs by rethinking music tokenization around a measurable notion of music semantic content grounded in downstream Music Information Retrieval tasks.
Guided by this definition, we propose \textbf{MuSeC}, a music semantic codec that factorizes semantic and acoustic content directly from mixed signals without source separation. MuSeC preserves information required for high-fidelity reconstruction while producing more LM-friendly discrete units. Empirically, it improves reconstruction quality and yields more predictable token sequences, providing a practical foundation toward high-fidelity LLM music generation. Demos are available at \url{https://longwaytog0.github.io/MuSeC/}.
\end{abstract}


\section{Introduction}
Recent progress in speech and music generation has increasingly converged on a three-stage language-model (LM) pipeline of discretization, modeling, and synthesis: a tokenizer maps waveform audio to discrete units, an autoregressive LM models the resulting sequences, and a vocoder or diffusion-based decoder renders the final waveform. As LM capacity and training practice continue to mature, generation quality is increasingly bounded by the tokenization stage, since the discrete representation jointly determines the LM's predictability and the fidelity attainable after decoding.

A central difficulty in audio tokenization is balancing reconstruction fidelity with LM predictability~\cite{zhang2023speechtokenizer, yang2023uniaudio}. Conventional neural codecs~\cite{defossez2022encodec, kumar2023dac, zeghidour2021soundstream} optimize perceptual reconstruction and tend to produce high-entropy token sequences that faithfully capture fine-grained acoustic detail but are difficult for an LM to model. Semantic-guided tokenizers mitigate this mismatch by aligning discrete units with higher-level content---e.g., through auxiliary automatic speech recognition objectives or constraints derived from self-supervised learning (SSL) representations---which reduces LM perplexity while retaining reconstruction quality~\cite{deng2025codecbench, wang2025audiocodecbench}. Extending this recipe to music is markedly harder: music exhibits richer polyphony, denser rhythmic organization, and broader timbral diversity, and vocals are typically entangled with accompaniment within a single mixture. This not only complicates the modeling of mixed signals but, more fundamentally, obscures what should count as musically meaningful content in the first place.

Prior efforts to bring semantic-guided tokenization to music~\cite{yuan2025yue, lei2025levo, lin2025duotok} largely inherit speech-centric notions of semantics. In doing so, they either retain speech-derived constraints that do not adequately capture musical structure, shift the tokenizer's role from reconstruction toward generation, or rely on explicit stem separation that incurs irreversible information loss in the mixture. These limitations suggest that progress requires \emph{rethinking} music tokenization itself, around two questions. \emph{First}, what constitutes \emph{music semantic content} in a way that is well-defined and measurable, rather than inherited from speech-centric assumptions. \emph{Second}, can the semantic and acoustic content of music be factorized within a single mixed signal---without source separation---while preserving high-fidelity reconstruction and yielding discrete units amenable to LM modeling.

We address the first question by grounding music semantic content in \emph{measurable musical competencies}: we operationalize it through the performance of music SSL representations on a suite of Music Information Retrieval (MIR) tasks~\cite{muller2015mir, yuan2023marble}, and use layer-wise probing to locate semantic-rich layers within SSL backbones~\cite{li2023mert, zhu2025muq}. Building on this definition, we introduce \textbf{MuSeC} (\textbf{Mu}sic \textbf{Se}mantic \textbf{C}odec), a reconstruction-oriented codec that factorizes semantic and acoustic components without source separation. MuSeC treats SSL-derived semantic content as a structural backbone for music and reconstructs the waveform by complementing it with an acoustic token stream that captures acoustic detail, achieving high reconstruction quality while yielding discrete units that are more favorable for LM modeling.

Our contributions are threefold. \emph{First}, we propose a measurable, MIR-grounded definition of music semantic content, together with a probing-based procedure for identifying semantic-rich SSL layers, moving beyond speech-inherited assumptions. \emph{Second}, we introduce MuSeC, a single-model music semantic codec that factorizes the semantic and acoustic content of \emph{mixed} music without source separation, while remaining focused on quantization and reconstruction. \emph{Third}, we provide empirical evidence that this factorization relaxes the usual tension between reconstruction fidelity and LM predictability, improving reconstruction quality while producing more LM-friendly tokens. We will also release training code and model weights to support reproducibility.

\section{Related Work}\label{sec:related}

\textbf{Neural Audio Codecs}\quad
Neural audio codecs compress continuous waveforms into discrete acoustic tokens while reconstructing high-fidelity audio, and most follow the VQ-GAN~\cite{vqgan} paradigm, in which a VQ-VAE-style~\cite{vqvae} encoder--quantizer--decoder is trained together with adversarial discriminators for perceptual quality. SoundStream~\cite{zeghidour2021soundstream} introduced residual vector quantization (RVQ) to improve quantization efficiency, EnCodec~\cite{defossez2022encodec} added recurrent layers to strengthen compression, and DAC~\cite{kumar2023dac} further improved fidelity through better quantizer and discriminator design. A parallel line of single-codebook codecs, including WavTokenizer~\cite{wavtokenizer} and BigCodec~\cite{bigcodec}, replaces multi-layer quantizers with a single codebook to reach ultra-low bitrates and simplify downstream modeling. Optimized primarily for reconstruction, however, these codecs lack explicit semantic constraints and thus produce high-entropy token streams whose fine acoustic detail is hard for an autoregressive LM to model~\cite{deng2025codecbench, wang2025audiocodecbench}.

\textbf{Codec with Semantic Supervision}\quad
To make tokens more predictable, a growing body of work injects semantic structure into the discrete units, differing mainly in how semantics enter the codec. SpeechTokenizer~\cite{zhang2023speechtokenizer} distills SSL features (from HuBERT~\cite{hsu2021hubert}) into the first RVQ level, and Mimi~\cite{defossez2024moshi} similarly distills WavLM~\cite{chen2022wavlm} features into a dedicated semantic quantizer, so that the leading tokens carry semantics while later levels refine acoustic detail; CosyVoice~\cite{du2024cosyvoice} instead derives supervised semantic tokens from an ASR encoder. Rather than distillation, SemantiCodec~\cite{liu2024semanticodec}, X-Codec~\cite{ye2025xcodec}, and XY-Tokenizer~\cite{gong2025xytokenizer} explicitly inject or fuse pre-trained semantic features with acoustic embeddings before quantization, and recent benchmarks confirm that such constraints lower LM perplexity while retaining reconstruction quality~\cite{deng2025codecbench, wang2025audiocodecbench}. These methods, however, are developed for speech, where semantic content is largely captured by phonetic or ASR targets, which may not adequately capture what is semantically salient in music.

\textbf{Music Codecs}\quad
Adapting these ideas to music has followed three routes. (i)~\emph{Retraining} a speech-oriented codec on music data, as in YuE~\cite{yuan2025yue}; because the semantic constraint still derives from speech-centric SSL objectives~\cite{chen2022wavlm, hsu2021hubert}, it may not capture musical structure in complex mixtures. (ii)~\emph{Quantize-then-generate}, as in LeVo~\cite{lei2025levo} and HeartMuLa~\cite{yang2026heartmula}, which directly quantize music SSL representations---such as those from MERT~\cite{li2023mert} and MuQ~\cite{zhu2025muq}---and delegate waveform rendering to a diffusion decoder; this eases the LM's modeling burden but shifts the tokenizer from quantization-and-reconstruction toward quantization-and-generation. (iii)~\emph{Stem-wise} tokenization, as in DUO-TOK~\cite{lin2025duotok}, which treats vocals as the semantic stream and accompaniment as the acoustic stream; the explicit decomposition attains low perplexity but degrades reconstruction, consistent with the irreversible information loss incurred when separating tightly coupled sources. MuSeC departs from all three: it retains a reconstruction-oriented objective on the \emph{mixed} signal and factorizes semantic and acoustic content without any source separation.

\begin{figure}[tb]
  \centering
  \includegraphics[width=1.0\linewidth]{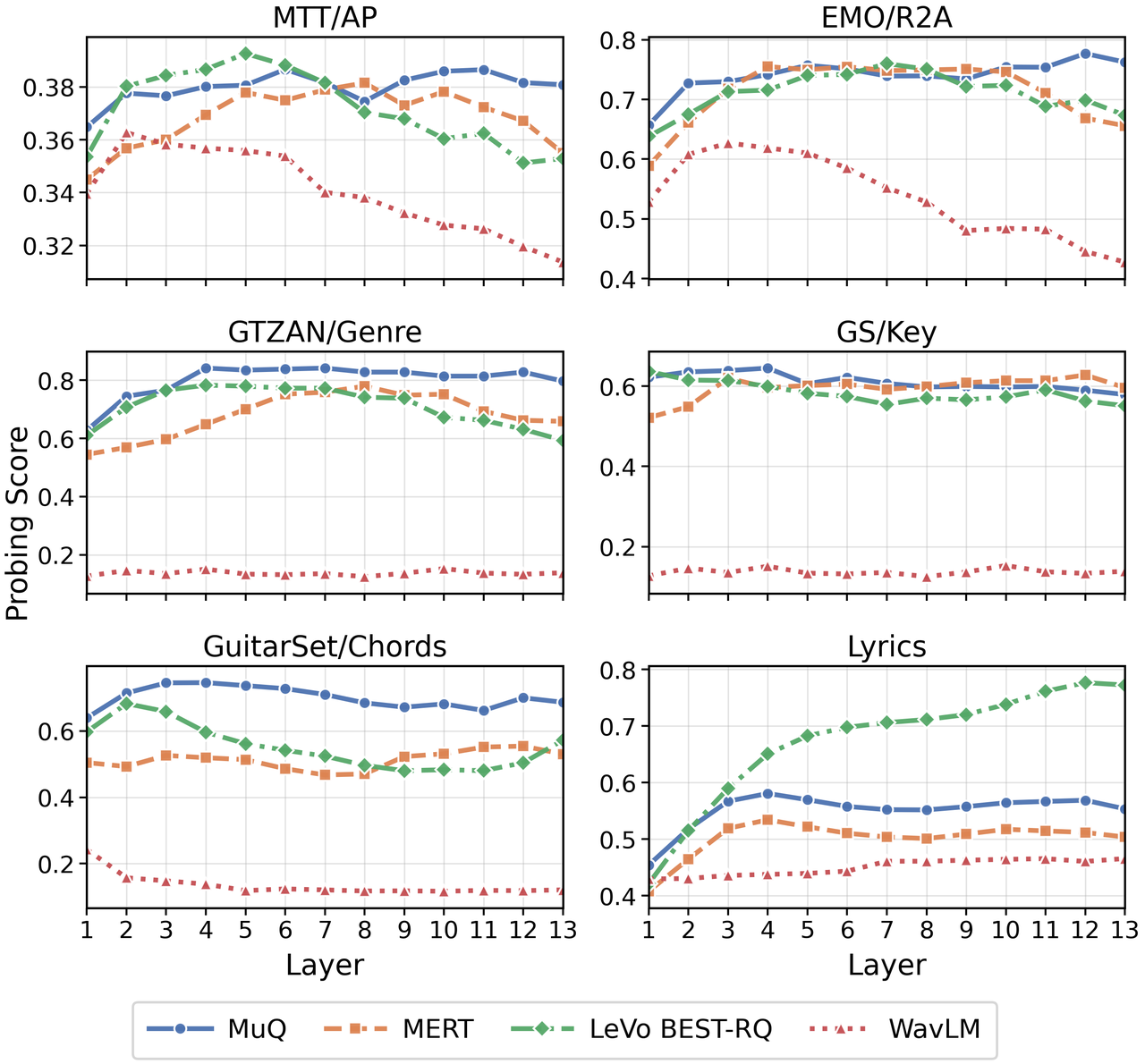}
  \caption{Layer-wise evaluation of SSL representations on different MARBLE tasks. WavLM is included as a speech-SSL reference to illustrate the speech--music semantic gap.}
  \label{fig:ssl-probe}
  \vspace{-10pt}
\end{figure}

\section{MuSeC}
\subsection{Music Semantic Content}
\label{sec:music_semantic}
Music is inherently more complex than speech. Its vocal component exhibits richer prosody and rhythmic patterns that convey nuanced affect, while accompaniment introduces additional musical structure such as harmony and key, giving rise to diverse genres and styles. Moreover, vocals and accompaniment are tightly coupled in real music. Therefore, defining \emph{music semantic content} must go beyond speech-like vocal content and account for musically salient information embedded in both accompaniment and vocal--accompaniment interactions. 

Layer-wise analyses of speech SSL models such as wav2vec 2.0~\cite{baevski2020wav2vec2}, HuBERT~\cite{hsu2021hubert}, and WavLM~\cite{chen2022wavlm} offer a useful starting point: they reveal a consistent hierarchy in which shallow layers tend to encode low-level acoustic properties, whereas deeper layers capture more semantic, content-related information. Motivated by this, we conduct a layer-wise probing study of music SSL representations to examine whether music exhibits a similar hierarchy and, if so, what constitutes its \emph{semantic} and \emph{acoustic} content.

\begin{figure*}[tb]
  \centering
  \includegraphics[width=0.85\linewidth]{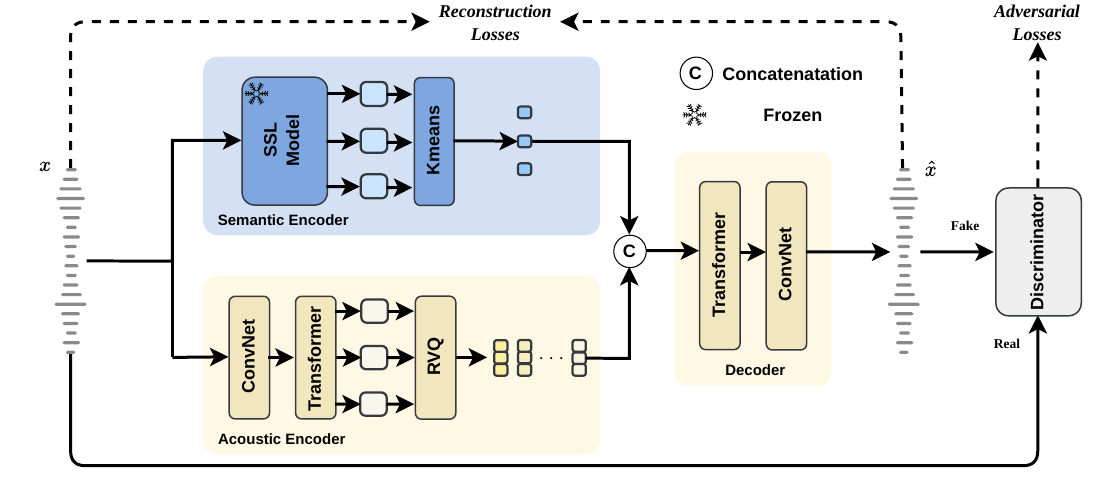} 
  \caption{Overview of MuSeC. A frozen SSL encoder + k-means provides semantic tokens; an acoustic RVQ stream captures fine-grained acoustics. The two streams are concatenated and decoded for high-fidelity reconstruction.}
  \label{fig:model}
  \vspace{-12pt}
\end{figure*}

Given an SSL encoder with $L$ layers, we extract layer-wise features $\mathbf{z}_{\ell}$ from each layer $\ell \in \{1,\ldots,L\}$ and train lightweight task-specific predictors on top of frozen features. The resulting layer-wise performance trajectories indicate where different musical attributes emerge along the representation hierarchy. We evaluate three music-oriented SSL backbones: MERT~\cite{li2023mert}, MuQ~\cite{zhu2025muq}, and the BEST-RQ version of MuEncoder used in LeVo. To further investigate the differences between music semantic content and speech semantic content, we additionally include WavLM~\cite{chen2022wavlm} as a speech-SSL reference.

Since our focus is on tokenization for music generation rather than music understanding, we concentrate on score-level and high-level description tasks and omit performance-level and acoustic-level tasks in MARBLE. Specifically, we evaluate music tagging (MTT/AP), genre classification (GTZAN/Genre), emotion recognition (EMO/R2A), key detection (GS/Key), beat tracking (GTZAN/Beat), chord estimation (GuitarSet/Chords), and lyrics transcription (Lyrics). Figure~\ref{fig:ssl-probe} summarizes the evaluation results; due to space, GTZAN/Beat is omitted from the figure, and it exhibits a trend consistent with GS/Key and GuitarSet/Chords.

The evaluation results reveal a clear stratification of music SSL models. Most MIR tasks exhibit a unimodal layer-wise profile, in which performance improves from shallow to middle layers and then converges or declines, indicating a progression from local signal properties to more abstract musical factors. Score-related attributes, such as key and beat, often peak earlier, consistent with their reliance on local regularities and their closer alignment with the structural scaffold of the signal. In contrast, high-level description tasks, including tagging, genre, and emotion, tend to peak in middle-to-deep layers, suggesting that these layers encode more global descriptors. Lyrics transcription behaves differently, since it is dominated by speech semantics and exhibits a distinct depth trend, with performance continuing to improve toward deeper layers rather than peaking in the middle layers as in most music-oriented tasks. In comparison, the speech-SSL reference WavLM consistently lags behind the music SSL models across all tasks in Fig.~\ref{fig:ssl-probe}, suggesting limited generalization from speech-pretrained representations to music and underscoring the need for music-specific semantic representations. This stratification answers our two questions: music does exhibit a speech-like hierarchy, and it grounds a measurable definition of musical content. We take the shallow-peaking score attributes (key, beat, and chord) as \emph{music acoustic content}, which forms the structural scaffold of the signal, and the deeper-peaking global descriptors (tagging, genre, and emotion) together with lyrics as \emph{music semantic content}.

Guided by this definition, we select the backbone that best encodes music semantic content. Since generation is conditioned on tags and lyrics, and LeVo BEST-RQ leads on both tagging-related tasks and lyrics transcription, we adopt \textbf{LeVo BEST-RQ} as the semantic backbone for MuSeC.

\begin{table*}[t]
  \caption{Main results. Arrows indicate whether higher ($\uparrow$) or lower ($\downarrow$) is better. Best results are in \textbf{bold}.}
  \label{tab:main_results}
  \centering
  \renewcommand{\arraystretch}{1.25}
  \small
  \resizebox{\textwidth}{!}{%
  \begin{tabular}{l c c cc c c ccc c ccc}
    \toprule
    \multirow{2}{*}{\textbf{Codec}} &
    \multirow{2}{*}{\makecell[c]{\textbf{Token}\\\textbf{Rate}}} &
    \multirow{2}{*}{\makecell[c]{\textbf{Codebook}\\\textbf{Size}}} &
    \multicolumn{2}{c}{\textbf{MTT}} &
    \multicolumn{1}{c}{\textbf{Chords}} &
    \multicolumn{1}{c}{\textbf{Lyrics}} &
    \multicolumn{3}{c}{\textbf{LM Top-$k$ Acc}} &
    \multicolumn{1}{c}{\textbf{LM}} &
    \multicolumn{3}{c}{\textbf{Reconstruction}} \\
    \cmidrule(lr){4-5}\cmidrule(lr){6-6}\cmidrule(lr){7-7}\cmidrule(lr){8-10}\cmidrule(lr){11-11}\cmidrule(lr){12-14}
    & & &
    \makecell[c]{AP$\uparrow$} &
    \makecell[c]{AUC$\uparrow$} &
    \makecell[c]{Acc$\uparrow$} &
    \makecell[c]{Score$\uparrow$} &
    \makecell[c]{Top-1$\uparrow$} &
    \makecell[c]{Top-5$\uparrow$} &
    \makecell[c]{Top-10$\uparrow$} &
    \makecell[c]{PPL$\downarrow$} &
    \makecell[c]{PESQ$\uparrow$} &
    \makecell[c]{STOI$\uparrow$} &
    \makecell[c]{Mel $L_{1}$$\downarrow$} \\
    \midrule

    XCodec-YuE &
    \makecell[c]{50} &
    \makecell[c]{8$\times$1024} &
    \textbf{0.347} & 0.884 & 0.348 & 0.352 &
    0.498 & 0.783 & 0.882 & 6.884 &
    1.642 & 0.617 & 1.905 \\

    MuCodec-LeVo &
    \makecell[c]{25} &
    \makecell[c]{1$\times$16385} &
    0.294 & 0.865 & 0.264 & 0.387 &
    0.391 & 0.577 & 0.662 & \textbf{1.575} &
    1.167 & 0.375 & 1.419 \\

    MuSeC &
    \makecell[c]{25} &
    \makecell[c]{1$\times$2048 + 8$\times$1024} &
    0.344 & 0.892 & 0.298 & 0.425 &
    0.628 & 0.889 & 0.932 & 1.613 &
    1.714 & 0.621 & 0.984 \\

    MuSeC &
    \makecell[c]{25} &
    \makecell[c]{1$\times$2048 + 16$\times$1024} &
    0.346 & \textbf{0.894} & \textbf{0.365} & \textbf{0.490} &
    \textbf{0.651} & \textbf{0.905} & \textbf{0.955} & 1.694 &
    \textbf{2.201} & \textbf{0.683} & \textbf{0.883} \\

    \bottomrule
  \end{tabular}%
  }
  \vspace{-10pt}
\end{table*}

\subsection{Architecture}
As shown in Fig.~\ref{fig:model}, MuSeC consists of a \emph{semantic encoder}, an \emph{acoustic encoder--decoder} with residual quantization, and an adversarial discriminator suite. Both streams operate at a frame rate of 25\,Hz, so their tokens are frame-synchronous by construction.

\textbf{Semantic Encoder}\quad
Inspired by LongCat-Audio-Codec~\cite{zhao2025longcatcodec}, the semantic encoder uses a frozen SSL model followed by a $k$-means quantizer that maps layer-wise SSL representations into discrete semantic units. Guided by the probing analysis in Sec.~\ref{sec:music_semantic}, which localizes semantic-rich representations to the intermediate layers of LeVo BEST-RQ, we restrict the search to candidate layers $\{3,6,9\}$ and cross-validate them against cluster sizes $\{1024,2048,4096\}$, selecting the configuration that best reconstructs audio from the quantized features alone. This yields SSL layer~6 with 2048 clusters, producing one semantic token per 25\,Hz frame.

\textbf{Acoustic Encoder and Decoder}\quad
The acoustic encoder consists of a lightweight \emph{causal} convolutional front-end, a Transformer backbone, and a Residual Vector Quantization (RVQ) module. Following Mimi~\cite{defossez2024moshi}, the causal convolutional stem performs streaming-compatible downsampling by a factor of 960, mapping the 24\,kHz stereo waveform to frame-level latents at 25\,Hz. We use a \emph{sliding-window Transformer} to model temporal dependencies with bounded complexity on long music sequences. The decoder mirrors this design with a sliding-window Transformer and a causal convolutional upsampling network. Because the acoustic stream shares the 25\,Hz frame rate of the semantic stream, the two are frame-synchronous, and their quantized features are simply concatenated channel-wise, without any resampling or alignment, before being decoded into the reconstructed waveform $\hat{x}$.

\textbf{Residual Vector Quantization}\quad
We use RVQ for the acoustic stream with codebook size 1024, 32 residual codebooks, and embedding dimension 1024. During training, we apply stochastic codebook dropout: we sample a truncation depth $m \sim \mathcal{U}\{9,\dots,32\}$ from the last three quarters of the codebooks and drop all subsequent ones, encouraging information to be compressed into the earlier stages so that the acoustic stream degrades gracefully under codebook truncation. At inference time, we retain the first 16 codebooks to form the acoustic tokens. MuSeC thus emits one semantic and 16 acoustic tokens per 25\,Hz frame.

\textbf{Discriminator}\quad
For adversarial training, we use a DAC-style discriminator~\cite{kumar2023dac} (MPD + MRD) together with a multi-scale subband CQT discriminator (MSS-CQT)~\cite{gu2024msscqtd}. Their adversarial and feature-matching losses are combined with reconstruction losses to improve perceptual quality.

\subsection{Multi-stage training}
\label{sec:multi_stage_training}
MuSeC is trained in three stages that decouple codebook construction, token learning, and final rendering.

\textbf{Stage~1: Semantic Codebook}\quad
We fit $k$-means on frozen SSL features to build the semantic codebook, done offline so that the acoustic stream has a stable, fixed semantic vocabulary to build upon.

\textbf{Stage~2: Codec Training}\quad
We train the full codec with the semantic tokenizer fixed, learning the acoustic encoder, RVQ, and decoder for waveform reconstruction. Keeping the semantic tokenizer frozen forces the acoustic RVQ to encode only the information absent from the semantic stream, which drives the factorization of semantic and acoustic content without an explicit disentanglement objective.

\textbf{Stage~3: Decoder Super-Resolution}\quad
We freeze the encoders and codebooks and retrain only the decoder on high-quality 48\,kHz audio, filtered by AudioBox~\cite{vyas2023audiobox} and SongEval~\cite{yao2025songeval}. Doubling the upsampling ratio, the decoder renders the fixed 25\,Hz tokens at 48\,kHz: the discrete representation stays fixed while the decoder specializes in high-fidelity, high-sample-rate synthesis.

\subsection{Training objective}
\label{sec:training_objective}
MuSeC is trained with reconstruction and adversarial objectives. Let $x$ be the ground-truth waveform and $\hat{x}$ the reconstruction. The \textbf{total generator loss} is
\begin{align}
\mathcal{L}_{G}
&=
\lambda_{\text{stft}}\mathcal{L}_{\text{MR-STFT}}(x,\hat{x})
+\lambda_{\text{stereo}}\mathcal{L}_{\text{SD-STFT}}(x,\hat{x})
+\lambda_{\text{c}}\mathcal{L}_{\text{commit}}
\nonumber\\
&\quad+
\lambda_{\text{adv}}\mathcal{L}_{\text{adv}}(\hat{x})
+\lambda_{\text{fm}}\mathcal{L}_{\text{fm}}(x,\hat{x}).
\end{align}
Here, $\mathcal{L}_{\text{MR-STFT}}$ is the multi-resolution STFT loss and $\mathcal{L}_{\text{commit}}$ is the RVQ commitment loss; the RVQ codebooks are updated via exponential moving average (EMA), while $\mathcal{L}_{\text{commit}}$ keeps the encoder outputs close to the selected codewords. $\mathcal{L}_{\text{SD-STFT}}$ is a stereo sum-and-difference STFT loss computed on the sum (mid) and difference (side) channels, which encourages the decoder to preserve the stereo image rather than only the mono content, an aspect that matters for music. The adversarial term $\mathcal{L}_{\text{adv}}$ and the feature-matching term $\mathcal{L}_{\text{fm}}$ are computed over the two discriminators described above (DAC-style and MSS-CQT), with $\mathcal{L}_{\text{adv}}$ taking the least-squares form and $\mathcal{L}_{\text{fm}}$ the $L_1$ distance between their intermediate feature maps. Unless otherwise noted, all loss weights are set to $1$ and were not tuned, indicating that MuSeC does not rely on careful loss balancing.

\begin{table*}[t]
  \caption{Ablation results. \textbf{S1A16} uses 1 semantic codebook and 16 acoustic codebooks; \textbf{S0A16} keeps the semantic branch but disables semantic tokens; \textbf{A16} removes the semantic branch. \textbf{LeVo BEST-RQ (Kmeans)} denotes the selected SSL layer after $k$-means quantization and is reported as a \emph{semantic reference}.}
  \label{tab:ablations}
  \centering
  \renewcommand{\arraystretch}{1.15}
  \footnotesize
  \resizebox{\textwidth}{!}{%
  \begin{tabular}{l cc c c ccc c ccc}
    \toprule
    \multirow{2}{*}{\textbf{Codec}} &
    \multicolumn{2}{c}{\textbf{MTT}} &
    \multicolumn{1}{c}{\textbf{Chords}} &
    \multicolumn{1}{c}{\textbf{Lyrics}} &
    \multicolumn{3}{c}{\textbf{LM Top-$k$ Acc}} &
    \multicolumn{1}{c}{\textbf{LM}} &
    \multicolumn{3}{c}{\textbf{Reconstruction}} \\
    \cmidrule(lr){2-3}\cmidrule(lr){4-4}\cmidrule(lr){5-5}\cmidrule(lr){6-8}\cmidrule(lr){9-9}\cmidrule(lr){10-12}
    & \makecell[c]{AP$\uparrow$} & \makecell[c]{AUC$\uparrow$} &
      \makecell[c]{Acc$\uparrow$} &
      \makecell[c]{Score$\uparrow$} &
      \makecell[c]{Top-1$\uparrow$} & \makecell[c]{Top-5$\uparrow$} & \makecell[c]{Top-10$\uparrow$} &
      \makecell[c]{PPL$\downarrow$} &
      \makecell[c]{PESQ$\uparrow$} & \makecell[c]{STOI$\uparrow$} & \makecell[c]{Mel $L_{1}\downarrow$} \\
    \midrule

    \makecell[l]{LeVo BEST-RQ\\(Kmeans)} & 0.345 & 0.891 & 0.171 & \textbf{0.499} &
                   -- & -- & -- & -- &
                   -- & -- & -- \\

    \midrule

    MuSeC-S1A16 & \textbf{0.346} & \textbf{0.894} & \textbf{0.365} & 0.490 &
                 \textbf{0.651} & \textbf{0.905} & \textbf{0.955} & \textbf{1.694} &
                 2.201 & 0.683 & 0.883 \\

    MuSeC-S0A16 & 0.292 & 0.854 & 0.358 & 0.443 &
                 0.532 & 0.807 & 0.882 & 5.771 &
                 1.540 & 0.616 & 1.178 \\

    MuSeC-A16   & 0.286 & 0.850 & 0.360 & 0.317 &
                 0.582 & 0.811 & 0.902 & 6.543 &
                 \textbf{2.246} & \textbf{0.689} & \textbf{0.831} \\
    \bottomrule
  \end{tabular}%
  }
\end{table*}

\section{Experiments}

\subsection{Training details}
\label{sec:exp_training_details}
\hspace*{\parindent}\textbf{Datasets}\quad
Training data are drawn from large-scale public music corpora. Stage~1 uses 5,000 hours of music, segmented into $10$\,s clips, to extract SSL features for $k$-means training. Stage~2 uses 120,000 hours of music at $24$\,kHz, randomly cropped into $6$\,s segments to train the codec. Stage~3 uses 20,000 hours of high-quality music at $48$\,kHz. For lyrics transcription evaluation, since the corresponding MARBLE datasets have limited Chinese/English coverage, we additionally construct a small bilingual test set that is held out from the training pool and excluded from codec training.

\textbf{Training Configuration}\quad
We use AdamW ($\beta_1=0.5$, $\beta_2=0.9$) to optimize the generator and discriminators, with an initial learning rate of $2\times10^{-4}$ under cosine decay with linear warmup. A weight decay of $5\times10^{-2}$ is applied to the Transformer backbones and disabled for all other parameters. The frozen semantic encoder has 330M parameters and is not updated during training; the trainable acoustic encoder--decoder comprises a 45M ConvNet and a 135M Transformer backbone. Stage~1 fits the $k$-means quantizer on CPU for 6 epochs. Stages~2--3 are trained on 8 NVIDIA A100 80GB GPUs with a per-GPU batch size of 2 (total batch size 16); Stage~2 trains for $800$k steps and Stage~3 for $1$M steps.

\subsection{Evaluation tasks and metrics}
\label{sec:eval_metrics}

We evaluate MuSeC along three axes, namely \emph{content preservation}, \emph{LM friendliness}, and \emph{reconstruction quality}. Content preservation covers music semantic content (including its linguistic/lyrics component) and music acoustic content.

\textbf{Content Preservation}\quad
Following the two-way distinction established in Sec.~\ref{sec:music_semantic}, we probe how well the tokens preserve music semantic content and music acoustic content. For music semantic content, we evaluate music tagging on MagnaTagATune (MTT), the standard benchmark for high-level music description, together with lyrics transcription for its linguistic component, using a lightweight CTC-based probe on the frozen tokens and reporting a normalized score derived from its CTC loss (higher is better). For music acoustic content, we evaluate chord estimation on GuitarSet, a standard chord benchmark that reflects the local, score-level structure targeted by the acoustic stream. These tasks serve as representative single-task proxies for the broader trends established by the layer-wise probing in Sec.~\ref{sec:music_semantic}.

\textbf{LM Friendliness}\quad
We adopt a multi-token prediction (MTP) setup similar to Marvis-TTS\footnote{\url{https://github.com/Marvis-Labs/marvis-tts}}: a 500M-parameter autoregressive LM predicts the first codebook, while an MTP module predicts the remaining codebooks conditioned on the LM outputs. We report perplexity (PPL) and Top-$k$ token accuracy ($k\in\{1,5,10\}$) on the first-codebook predictions.

\textbf{Reconstruction Quality}\quad
We report PESQ~\cite{rix2001pesq}, STOI~\cite{taal2010stoi}, and the Mel-spectrogram $L_1$ distance between reference and reconstructed audio. As PESQ and STOI are defined for single-channel speech, we compute them per channel on the left and right channels and average the two, using them as perceptual-quality and intelligibility proxies; the Mel-spectrogram $L_1$ distance complements them by capturing full-band spectral fidelity that these speech-oriented metrics do not cover.

\subsection{Evaluation Results}
\label{sec:eval_results}
Table~\ref{tab:main_results} presents results for the three evaluation axes in Sec.~\ref{sec:eval_metrics}. We benchmark two MuSeC configurations, each with one semantic codebook and either 8 (reduced) or 16 (full) acoustic codebooks, to enable a more direct comparison with XCodec-YuE and to illustrate acoustic-stream scaling. Overall, MuSeC delivers the best reconstruction quality and the highest Top-$k$ accuracy while remaining competitive on every content-preservation metric, achieving the strongest overall trade-off among the three axes.

\textbf{Content preservation}\quad
The full MuSeC configuration achieves the best overall content preservation, with the strongest results on music semantic content (MTT AUC and lyrics) and music acoustic content (chords). It is essentially on par on MTT AP, where the gap to XCodec-YuE (0.347 vs.\ 0.346) is within noise. Notably, semantic content (MTT and lyrics) is preserved comparably by both configurations, whereas chord accuracy improves markedly from the reduced to the full setting. This indicates that additional acoustic codebooks mainly benefit music-acoustic structure and leave the semantic content intact; we examine this complementary behavior further in Sec.~\ref{sec:ablations}.

\textbf{LM friendliness}\quad
Both MuSeC configurations substantially improve Top-$k$ accuracy over XCodec-YuE despite operating at half the frame rate (25 vs.\ 50\,Hz), strengthening the claim that semantic--acoustic factorization yields a more structured token space than simply increasing token density. MuCodec-LeVo attains the lowest first-codebook PPL, but this reflects its single semantic codebook, whose tokens are trivially predictable at the cost of severely degraded reconstruction and content preservation; its Top-$k$ accuracy remains well below both MuSeC and XCodec-YuE.

\textbf{Reconstruction quality}\quad
MuSeC consistently outperforms the baselines in reconstruction quality, with the largest gains in the full setting. Notably, even the reduced setting improves PESQ, STOI, and Mel-$L_1$ distance over XCodec-YuE while operating at half the frame rate, which also halves the sequence length the LM must model. Increasing the acoustic codebooks from 8 to 16 further improves perceptual quality and intelligibility, demonstrating that the acoustic stream scales reconstruction fidelity while preserving the semantic benefits of the codec design.

\begin{figure*}[tb]
  \centering
  \includegraphics[width=0.85\linewidth]{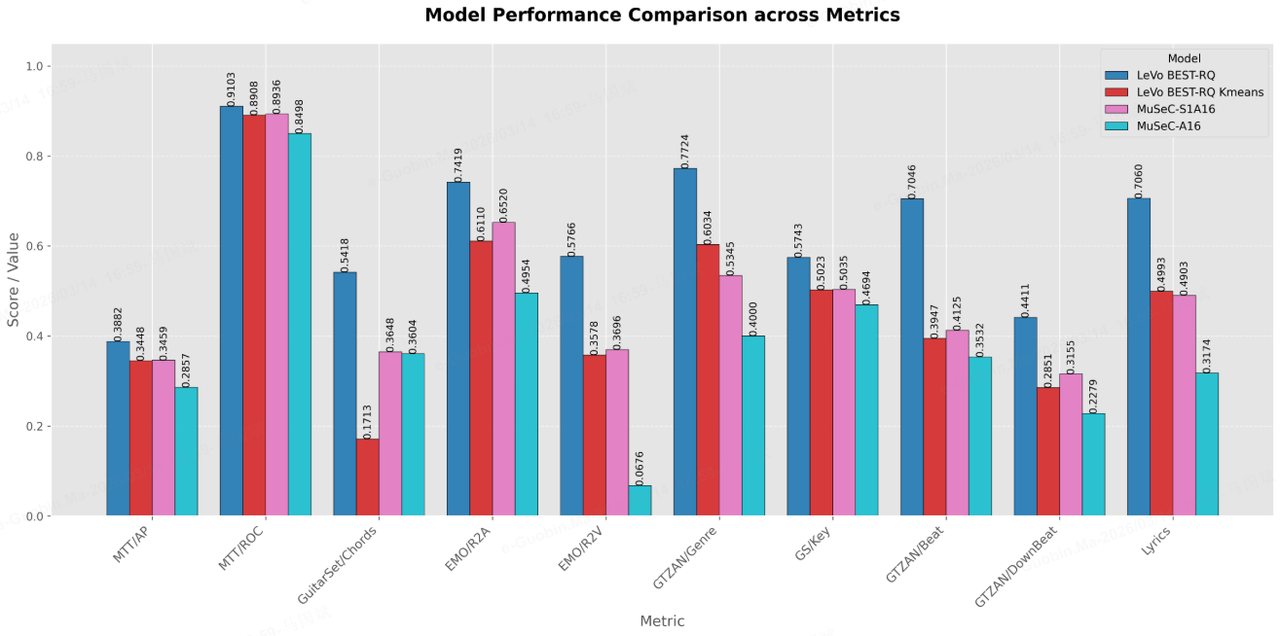} 
  \caption{MIR task performance for four settings: the continuous LeVo BEST-RQ representation (semantic topline), its $k$-means quantized form (semantic tokens only), the full MuSeC (S1A16), and the acoustic-only variant (A16). Evaluation follows Sec.~\ref{sec:music_semantic}.}
  \label{fig:ablation-probe}
\end{figure*}

\section{Ablations}
\label{sec:ablations}
Table~\ref{tab:ablations} highlights the role of the semantic codebook and its interaction with the acoustic stream. Disabling semantic tokens (S0A16) consistently degrades content preservation, LM friendliness, and reconstruction quality, indicating that the semantic codebook serves as a crucial backbone for both waveform reconstruction and autoregressive modeling. In contrast, removing the semantic branch (A16) yields the best reconstruction scores but substantially harms LM friendliness, suggesting that a purely acoustic RVQ representation can be optimized for fidelity while producing less predictable tokens. Notably, the full model (S1A16) closely matches the semantic reference (LeVo BEST-RQ after quantization) for both music semantic content (MTT) and lyrics, showing that semantic constraints are effectively preserved during quantization and reconstruction.

Compared with the original continuous LeVo BEST-RQ representation (Fig.~\ref{fig:ssl-probe}, layer~6), the $k$-means quantized semantic tokens alone lose substantial acoustic information: chord accuracy drops from $0.542$ to $0.171$, while music semantic content is largely retained. Interestingly, this missing acoustic content is largely recovered once the acoustic codebooks are introduced: chord accuracy rises back to $0.365$ in the full MuSeC (S1A16), at almost no cost to semantic content (e.g., MTT AUC $0.891\!\to\!0.894$, lyrics $0.499\!\to\!0.490$). We further investigate this complementary behavior across all MIR tasks. As shown in Fig.~\ref{fig:ablation-probe}, the two streams exhibit a clean double dissociation: adding the acoustic stream restores acoustic content (chord, key, beat) while leaving semantic scores intact, whereas adding the semantic stream restores semantic content while barely moving the acoustic scores, with lyrics improving from $0.317$ to $0.490$ and emotion (R2V) from $0.07$ to $0.37$. Notably, S1A16 and A16 reach almost identical chord accuracy ($0.365$ vs.\ $0.360$), confirming that the chord content in the full model is supplied almost entirely by the acoustic stream. Although it still trails the continuous topline ($0.542$), the acoustic stream already recovers much of the acoustic content lost to the $k$-means discretization of the semantic tokens ($0.171$).

This complementary behavior, in which each stream recovers only its own content type without disturbing the other, provides direct evidence that our task-grounded definitions of \emph{music semantic content} and \emph{music acoustic content} capture distinct factors, and that MuSeC's two-stream design effectively separates them.

\section{Conclusion}
\label{sec:conclusion}

In this study, we propose MuSeC, a reconstruction-oriented codec for high-fidelity LLM music generation. MuSeC uses a task-grounded definition of music semantics, obtained by evaluating SSL representations on MARBLE MIR tasks. It adopts a music-specific SSL backbone to provide semantic tokens. With a semantic-acoustic decoupled design operating on mixed music, MuSeC preserves high-level musical attributes and singing semantics, complemented by an acoustic RVQ stream for faithful waveform reconstruction. Our experiments demonstrate that MuSeC achieves stronger content preservation and higher reconstruction quality than existing tokenizers, and produces discrete units more amenable to autoregressive language modeling.

\end{document}